# ESCAPE

Preparing Forecasting Systems for the
Next generation of Supercomputers

# D2.2 Additional key features required for different directives based porting approaches

Dissemination Level: Public

This project has received funding from the European Union's Horizon 2020 research and innovation programme under grant agreement No 67162

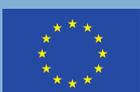 Funded by the European Union

Co-ordinated by

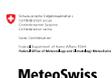 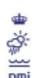 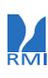 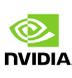 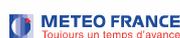 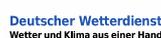 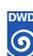 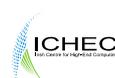 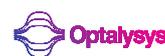 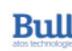 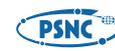 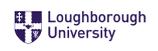

# ESCAPE

**Energy-efficient Scalable Algorithms for Weather Prediction at Exascale**

Author **Alastair McKinstry (ICHEC)**

Date **21/12/2017**

# Table of Contents

# Tables

# 1 Executive Summary

This report summarizes key features required for OpenMP and OpenACC directives based on experience in the ESCAPE project. For OpenMP, the latest public draft standard, 5.0, contains the deep copy and multi-level memory features desired; for OpenACC, Technical Report 16 summarizes ongoing discussions beyond the Standard version 2.6. This document includes a summary of our recommendations on this approach.

Additional work is also desirable in coordinating the runtime and debugging when both OpenACC and OpenMP directives are used; in particular the interoperability of the new OPDT debugging interface for OpenMP and its semantics in a mixed-directive program.

# 2 Introduction

## 2.1 Background

During the course of the ESCAPE project, dwarfs have been optimized using OpenACC and OpenMP directives for a number of systems and compilers. Typically, compilers have supported OpenMP 4.5 [1] (released November 2015) and OpenACC 2.5 [2] (released October 2015), though compiler versions before this have used older versions.

During the project, a number of draft standard changes have been proposed; OpenACC 2.6 [3] and OpenMP 5.0 [4].

## 2.2 Scope of this deliverable

### 2.2.1 Objectives of this deliverable

This deliverable reports recommendations to changes in the OpenMP and OpenACC standards based on issues encountered in ESCAPE.

### 2.2.2 Work performed in this deliverable

The changes recommended in this report are being forwarded as input to the OpenACC and OpenMP standards committees, specifically in response to OpenACC technical report 16 [5]

### 2.2.3 Deviations and counter measures

No deviations or countermeasures were required.

# 3 Data transfer between memory spaces

## 3.1 Deep copy support in OpenACC

In current releases, neither OpenMP nor OpenACC have deep copy support for hierarchical objects, though this has been a desirable feature much discussed in standards bodies.

Consider the `geometry_type` used within the Atlas library in ESCAPE. Here a number of integer and real arrays are defined, some of which (e.g. iedge2node) are needed on the device in high-bandwidth memory, others may be on the host alone or in shared memory.

```
type, public :: geometry_type
  integer, public :: nb_edges
  integer, public :: nb_nodes
  integer, public :: nb_levels
  type(atlas_connectivity)                 , public  :: edge2node
  type(atlas_connectivity)                 , public  :: node2edge
  real(wp), pointer                        , public  :: rpole_bc(:)
  real(dp), allocatable                    , public  :: node2edge_sign(:,:)
  integer,  pointer                        , public  :: is_pole_edge(:)
  integer,  pointer                        , public  :: pole_edges(:)
  integer,  pointer                        , public  :: iedge2node(:,:)
  integer                                  , public  :: nb_pole_edges
  real(wp), pointer                        , public  :: dual_volumes(:)
  real(wp), pointer                        , public  :: dual_normals(:,:)
  real(wp), pointer                        , public  :: lonlat(:,:)

  type(atlas_StructuredGrid)                         :: grid
  type(atlas_Mesh)                                   :: mesh
  type(atlas_functionspace_nodecolumns), public :: nodes
  type(atlas_functionspace_edgecolumns), public :: edges

  type(atlas_Field)                        , private :: field_pole_bc
  type(atlas_Field)                        , private :: field_lonlat
  type(atlas_Field)                        , private :: field_dual_volumes
  type(atlas_Field)                        , private :: field_dual_normals

  type(atlas_FieldSet), public :: topology_fields

end type

type(geometry_type) :: geometry
```

Currently, if we use a simple directive

```
!$ACC copy(geometry)
```

then accesses to the arrays iedge2node() on the device will fail. It takes examination of the code to place explicit additional copies to make the code run:

```
!$ACC copy(geometry, geometry%iedge2node)
```

OpenACC Technical report 16 describes discussions on deep copy support for complex objects, where array allocations occur within elements of a larger class. Currently this is a major source of failures when developing object-orientated codes with OpenACC: the parent object may be mapped to the device, the child objects are not. Within Fortran, it is possible for the compiler to specify the size and layout (shape) of arrays, making it possible for the compiler and runtime to ensure this does not occur: the question is when and whether this is an appropriate behavior. For C/C++ a set of shape directives are added.

Using the new mapper directives within OpenMP 5.0 we may implement this behavior using mapper declarations, such as:

```
! Declare the type and map to device
type (device_int_array) :: iedge2node
!$OMP declare mapper(device_int_array::iedge2node) map(to)
```

The `use_by_default` clause for mapper enables all variables of a given type to be subject to a given mapper. This enables a "policy" to be constructed enabling a form of deep copy to be implemented.

### 3.1.1 Data layout control

No methods of data layout control are currently proposed for OpenACC under TR-16, but Q1.1 describes three possibilities, and asks for motivating examples. The equivalent functionality can be specified within OpenMP 5.0 using memory allocators. Equivalent memory allocators may be defined within OpenACC.

### 3.1.2 Attach functionality

New operations, attach and detach were added to OpenACC 2.6, which implements the option of explicitly performing the attach operation as required. This attach operation has performance implications and hence avoiding unnecessary automatic attach operations is desirable. Similarly, explicitly declaring attach operations moves the onus to the programmer of avoiding situations where there is no clear semantics of the correct value for a pointer, such as where a pointer x is in shared memory but the allocated array on a device is in device-private memory. Here, overwriting the shared pointer would lose the host-pointer value, while not attaching would make the device array inaccessible from the base pointer. In this case, the programmer needs to handle the memory explicitly and automatic attach will likely fail, and so the explicit usage is specified in the OpenACC 2.6 standard.

Again, the question of automatic attach is moot for Q2.3, where bottom-up data structure creation is described.

Q2.4 asks what the effect of a detach operation should be on the device pointer, when the associated attachment counter has decremented to zero. Setting to NULL is one possibility, as is leaving the pointer stale. Draft 2.6 answers the question by setting the device pointer to the value from local memory. This is the preferred solution as it may either be tested to be a valid device pointer (equivalent to NULL where it matters) or possibly be a shared pointer, in which case code may continue. This allows for the case of moving data from shared to device-private memory for optimisation.

### 3.1.3 Deep copy directives

TR-16 proposes the concept of policies for defining behaviour with hierarchical data structures. In answer to Q3., we agree with the proposed default of a deep copy for all shaped structures. This provides program correctness (automatic validity of members, unlike the current shallow behaviour) that can still be over-ridden for performance where necessary; the task of adding a new policy to handle these cases explicitly is not onerous.

For C/C++, S. 3.1.2 defines a shape directive to provide functionality equivalent to Fortran's array descriptor behaviour. We agree this should be added.

Member selection for policies would then be handled according to include and exclude clauses. We propose no alternative name suggestions for Q3.3, and see no examples where both include and exclude operations can apply to the same shaped member (Q3.4).

For Q3.5, the default data movement behaviour should then be as proposed: when a hierarchical data structure appears in a data clause, each member is treated as appearing in the same clause unless overridden by an explicit policy.

For Q3.6, We propose no alternative name to create, even though it does not involve data motion, as it is currently defined and understood. Q3.7 describes the action of the `nocreate` clause as added to version 2.6, and we agree with this as-is.

Q3.8 asks if an object should be traversed if the parent is already present. The draft proposes that a deep copy is indeed deep, and the implementation should ensure that the entire requested structure is present on a device. We agree, as otherwise there could be breakage in behaviour if new elements are added to a hierarchical structure: a programmer would be required to inspect all code to ensure additional 'present' statements are added, which breaks object encapsulation. Again, the behaviour can be overridden by defining new policies where performance impact is significant.

Q3.9 asks what happens in the case of bottom-up deep copies: what if a child is present, but excluded by current policy: does the pointer get translated or not. As the default behaviour requires the programmer to explicitly declare the 'exclude' we would expect it to be not translated.

For Section 3.2.1, Q3.9 asks how complex a `shape-expression` will need to be. We expect that it will need to be able to involve whatever constants, members of this struct, class datatype and member functions that are in scope at the point of the directive. In the second case presented, `member-name@member-name`, where the named member is an offset from the sibling member, when the named member is translated, the translated member will have the same offset as the original value has from the original sibling member. Hence the translated value may not point to valid device data.

It is likely to be unavoidable that any sufficiently complete expression for shape-expression will lead to cases where the value is an invalid device address, especially if different memory spaces are used.

Q3.11 asks if there should be an explicit `policy(default)`, and a `policy(*)` for clauses that apply to all policies. While not required, we support these as they can make code clearer in some instances.

There was discussion of hierarchy of policies in Q3.12, where a child policy can add or modify the behaviour of a parent policy. We have no motivating examples to present, and suspect that this will require too much complexity for little benefit.

Again, we see no reason to add both exclude and include clauses in the same policy for a datatype, as asked in Q 3.13.

### 3.1.4 Direction Clauses

A subtle point is described in Q3.14. Here a case is presented of an object `s1` of type `t1_t`, which contains a member `m2`. Within the definition of `t1_t`, `m2` a policy specifies a direction `inout(m2)`, while in the parent `t2_t`, a policy `in(s1)` is defined. The proposal asks, which should apply to `m2`, and is the direction specified for `s1` a default or limitation?

We propose that the policy `inout(m2)` should apply. This is on the grounds that this then observes encapsulation, where the behaviour of a type is fully specified. In the case of mixing OpenMP and OpenACC, the memory allocator and location of a member may be specified within the definition of `t1_t` and `s1`; changing the type and location of `s1` from, for example, shared to device-private memory may break behaviour if the copying behaviour is specified in arbitrary locations elsewhere in the code, not in the definitions of `t1_t` and `s1`.

### 3.1.5 Open issues

TR-16 leaves open the issues of

- Polymorphic datatypes in Fortran
- Dynamic types in C++
- Virtual function members in C++.

No answers to these issues are presented here, except to note that any proposed changes to OpenACC deep copy and presence behaviour need to be reconciled with the memory hierarchy and mapper functionality added to OpenMP 5.0. In this case, attaching behaviour to datatype rather than member is preferred.

With deep copy support defined, either via OpenMP or OpenACC, the compiler can provide warnings and guidance in the case of partial and incomplete deep copies. For example, when accessing a component of an object that has not been copied, and no present statement is provided, warnings should be issued to the programmer.

## 3.2 Memory layout and allocators

One open question in TR-16 is the memory layout when copying. Experience in ESCAPE is that only some members of large objects such as fields are required on devices, but layout and memory type is important for performance. TR-16 allows for a syntax for partial deep copies to be used as default, which we recommend.

OpenMP 5.0 extends the standard by allowing new allocators. Section 2.7 specifies new default allocators that may be used: eg. Default, `large_capacity`, `low_latency`, `high_bandwidth` and `omp_cgroup_mem_alloc`, `omp_pteam_mem_alloc`, which allow for memory attached to a particular parallel team of threads or contention group. These can then be used wither with runtime `omp_alloc()` calls, or with allocator clauses in directives.

Interoperability between the directives standards would then require OpenACC to allow variables declared in this way to be automatically seen as present, and enable the specification of different components of large memory objects on different memory spaces. While at present there is no proposal for OpenACC functionality to match this memory allocator behavior in OpenMP, explicit interoperability, especially in the case of deep copy functionality being added to OpenACC, should be defined.

### 3.2.1 Compiler stability

Compilers are getting more and more stable and mature but we can still experience compiler crashes (internal error / segmentation faults). These errors are then very hard to localize and minimal examples reproducing them are hard to set up in order to issue a bug report to compiler vendors.

Newer Fortran constructs are the one that are most likely to fails (e.g. derived-types).

### 3.2.2 Difference between compilers

For some functionalities that are not specified in the standard, different compilers provide different behaviour in the implementation. The main example of this situation is the support of deep copies (previously mentioned). Cray is implementing an option for the deep copy, where PGI is not and doing the deep copy explicitly with directive for the type members results in different results between PGI and Cray. While the aforementioned attach functionality of the 2.6 standard would improve the support for deep copies, this version of the standard will not be implemented by all vendors, limiting though the portability of the GPU codes.

### 3.2.3 New features

Array notation in Fortran is not covered by the standard and therefore Fortran codes have to be refactored in order to be used with the parallel/loop directive. Since Fortran models make extensive use of the array notation, support for it in OpenACC would be of great value, as it keeps a high level of expressiveness in the code. Using the kernels directive does not guarantee that it will parallelize the statement. In particular if arrays in the vector notation are pointer, some aliasing might block this parallelization.

### 3.2.4 Scalar ambiguity in the standards

Using a scalar variable without any clause can be confusing. According to the standard: "a scalar variable referenced in the parallel construct that does not appear in a data clause for the construct or any enclosing data construct will be treated as if it appeared in a firstprivate clause". But when reading the specification for the firstprivate clause: "The firstprivate clause is allowed on the parallel construct; it declares that a copy of each item on the list will be created for each parallel gang".

Since a copy is created for each gang, can one assume that all threads in a gang are accessing the same copy? This would mean that the following code is potentially incorrect with a race condition on tmp (assuming this variable is only used in this statement here), since all thread in a gang would write to it.

```
!$acc parallel copy(a)
!$acc loop gang
DO k=1,nk
  !$acc loop vector
  DO i=1,ni
    tmp=2*i*k
    a(i,k)=tmp
  END DO
END DO
```

Another point that can be easily misunderstood by user, is the use of the same clause with different directives that have different behaviour. The private clause as a different behaviour whether it is used with the parallel directive or with the loop directive using the vector clause. This different behaviour can mainly be enforce by the compilers using data analysis. (Note: this is actually corrected in OpenACC 2.6).

### 3.2.5 Programming challenges coupling OpenACC with other approaches

In a big and complex setup like a large atmospheric model, the port can be a mix between OpenACC directive and re-write using Fortran and an accelerator specific language such as CUDA, C/CUDA or C++/CUDA. In this setup, the interoperability between the FORTRAN/OpenACC and the other language like CUDA code can be problematic. It is easy to allocate arrays on the FORTRAN/OpenACC side and then pass the pointer to the C/C++ code with CUDA. However allocating arrays with CUDA and passing the GPU memory pointers to OpenACC environment is not defined and supported by the standard.

Within the ESCAPE project this was implemented making use of the acc_map_data that is only available from C or C++. Although this has been demonstrated in the use of OpenACC versions of ESCAPE dwarfs, which allocate memory using the Atlas library in C++, there is no guarantee from the standard that the runtime lookup tables of the fields in the host and device memory spaces is shared between C++ and Fortran.

As a consequence the interoperability between C++/CUDA and Fortran OpenACC is not guarantee by the standard.

This issue has been reported to vendors (PGI/NVIDIA) within the ESCAPE project.

### 3.2.6 Additional compiler features

For debugging purpose, it would be very useful to provide support from compilers for a –O0 option that will be able to generate bit-reproducible results between a CPU compiled version and a GPU compiled version. Bugs due to wrong usage of the OpenACC data or parallelization directives, or due to compiler bugs, are common in production codes and increase considerably the maintenance cost of models that are ported to GPUs. Once a bug is detected, since the results of the simulations of CPU and GPU diverge beyond the tolerance error, debugging it becomes a daunting task as there is no easy way or tools to validate the results produced by a kernel or a parallel region or sections of it. All architectures guarantee precision of computations up to IEEE floating point precision. However the error of computations within parallel regions will propagate differently in different architectures. If results differ for CPU and GPU runs, it is difficult to evaluate if these differences are within expected tolerance errors or there are hidden bugs. A bit-reproducible mode between CPU and GPU implementations would facilitate the debugging task by ensuring that each parallel region should generate bit reproducible results.

### 3.3 OpenMP / OpenACC interoperability

Different compilers vary in their degree of interoperability for OpenMP and OpenACC.

GNU compilers allow combined OpenMP and OpenACC constructs, using the same runtime. OpenMP constructs may not be nested within OpenACC.

Cray adheres strictly to standards and is less flexible and combining OpenACC directives within OpenMP regions generates runtime errors. To date, only "omp target" directives are supported.

PGI allows flexibility in mixing OpenACC and OpenMP directives beyond the current standards.

Individual statements set syntax rules that make mixing complex: in particular, both the `$OMP PARALLEL DO` and `$ACC LOOP` directives require to be immediately prior to a stated DO loop.

It may be desirable to have both present and operable dependent on compiler flags provided; depending on any conditional clauses attached (eg. simd or vector clauses), immediately nested OMP and ACC loops may be redundant, or scheduled as vector parallelism within threaded parallelism, for example. Specification of when nested OMP and ACC loop directives are allowed should be clarified. It appears OpenACC loops directly within OpenMP loop directives should be considered legal, but not vice versa; simple OpenACC loops without vector clauses may be considered as redundant and flagged as such, but not prohibited.

### 3.3.1 Compiler support and optimisation

The long-term future of OpenACC and OpenMP inter-operability is unclear at this stage. Recently Cray have signalled their intention to drop support for OpenACC in favour of OpenMP in their compilers. With draft 5.0 features in OMP, it appears that all functionality, if not ease of programming, will be possible with OpenMP alone.

For compiler auto-parallelisation, greater clarity is required as to whether directives then overrule compiler optimisations. For example, when applying OpenMP threading to Fortran DO CONCURRENT loops, the GNU compiler typically will optimise the DO loop adding vector parallelism. Does the OpenMP directive then override this, or add a second layer of threaded parallelism?

### 3.3.2 Reduction operations in OpenACC

In reduction operations (OACC), reduction operations work on a scalar only. When operating across a team on a device, a private instance of a variable is created for each parallel gang. At the end of a region, the values for each gang are combined using the reduction operator and the result stored in the original variable.

As of OpenACC 2.6, support is not present for reduction into an array, or of an array by one dimension.

E.g in Fortran:

```
!$ Legal in OpenACC
!$ACC reduction(a:max)
a = max( res(:) )

! Not legal
!$ACC reduction(maxres:max)
maxres(:) = max(res(:, 1))
```

Cannot be optimised using OpenACC.

In OpenMP, reduction in dimension is possible, as here:

```
#pragma omp parallel for reduction(a:+) private(j) \
for (i = 0; i< N ;i ++)
    for (j = 0; j < M; j++)
        a[j] = b[i][j]
```

It is desirable that this is extended for OpenACC. Unfortunately as above, it is not possible to place the OpenMP construct within an enclosing OpenACC loop. Hence we propose that reduction in OpenACC be extended to vectors, matching OpenMP.

### 3.3.3 Loop optimisation

Loop (or loop nest) optimization represents a critical aspect of parallelizing code on both native or accelerated base systems as loops represent most of the time in the most computational part of an application.

Thus, providing a set of low level directives (compared to current approaches) allowing the developers to specify and perform more loop transformations as loop permutation, distribution, unrolling (full, partial, unroll-and-jam), etc. is desired.

In addition, giving more information to the compiler on the loop properties (number of iteration, that number of iterations is a multiple of X, number of iteration between Y and Z with a stride of S, etc.) can help it at determining the best approach to efficiently parallelizing the loop (nest) on a specific hardware target.

This approach is less portable across different platform but can help at better take efficiently advantage of the hardware capabilities.

Such directives have been proposed in the past by a source-to-source compiler to build parallel hybrid applications running on many-core systems, including GPU [6].

## 4 Debugging: OMPD

OpenMP 5.0 adds a new interface, OMPD, which enables third party support for debugging. This extends the existing first-party OPMT API. No such comprehensive interface exists for the OpenACC standard. Nevertheless, OMPD should enable tracing of mixed OpenMP / OpenACC codes (or OpenACC codes where OpenMP tracing has been enabled).

By design with OMPD, an OpenMP implementation provides a plugin supporting callbacks. The OMPD interface is tied directly to the OpenMP runtime state, as defined in e.g. 4.2.2.10 of TR-6. No strict mapping between OpenACC and OpenMP runtime states is currently defined, although in practice such mappings will exist in runtime environments.

Hence to support debugging of OpenACC programs, a mapping of OpenACC and OpenMP 5.0 OMPD states is highly desirable.

*Table 1: Compilers used, with directive support*

| Compiler | Version | Support |
|---|---|---|
| PGI | 17.1 | OpenCC 2.5, OpenMP 3.1 |
| Cray | 16.1 | OpenMP 3.1, OpenACC 2 ? |
| GNU | 7.2 | OpenMP 2.5, OMP 4.5 |
| Intel | 17.0 | OpenACC 2.5, OpenMP 4.5 |

## 5   Conclusion

Future directive support in compilers should include handing of directive-compatible handling of memory areas in accelerator based systems. Initial work on this is included in the OpenMP TR-4 report and we recommend its progress.

Additionally, deep copy and presence support is needed for complex codes such as the Atlas library used in ESCAPE. Again, draft report TR-16 for OpenACC accelerators describes initial work in this area, and we provide input for that report.

These two main features need to be provided in a compatible run-time, enabling deep copy and presence to be specified to specific memory areas, such as High-Bandwidth memory or other shared memory areas.

OpenMP 5.0 is currently in draft, with an expected publication date of November 2018. At this stage, minor clarifications such as the in S.3.3.1 will be raised. For OpenACC, deep copy support is unlikely in version 2.6 but the conclusions of this report will be forwarded to the TR-16 discussion for potential inclusion in any future OpenACC 2.7/3.0.

## Document History

| Version | Author(s) | Date | Changes |
|---|---|---|---|
| 0.1 | Alastair McKinstry | 05/12/2017 | Initial Version |
| 0.2 | Alastair McKinstry | 11/12/2017 | Included changes from MSWISS and BULL |
| 1.0 | Alastair McKinstry | 21/12/2017 | Final version after review |
| | | | |

## Internal Review History

| Internal Reviewers | Date | Comments |
|---|---|---|
| Erwan Raffin (Bull) | 19/12/2017 | Approved with comments |
| Alex McFadden (OSYS) | 19/12/2017 | Approved with comments |
| | | |
| | | |

## Effort Contributions per Partner

| Partner | Efforts |
|---|---|
| ICHEC | 1 |
| | |
| | |
| **Total** | 1 |

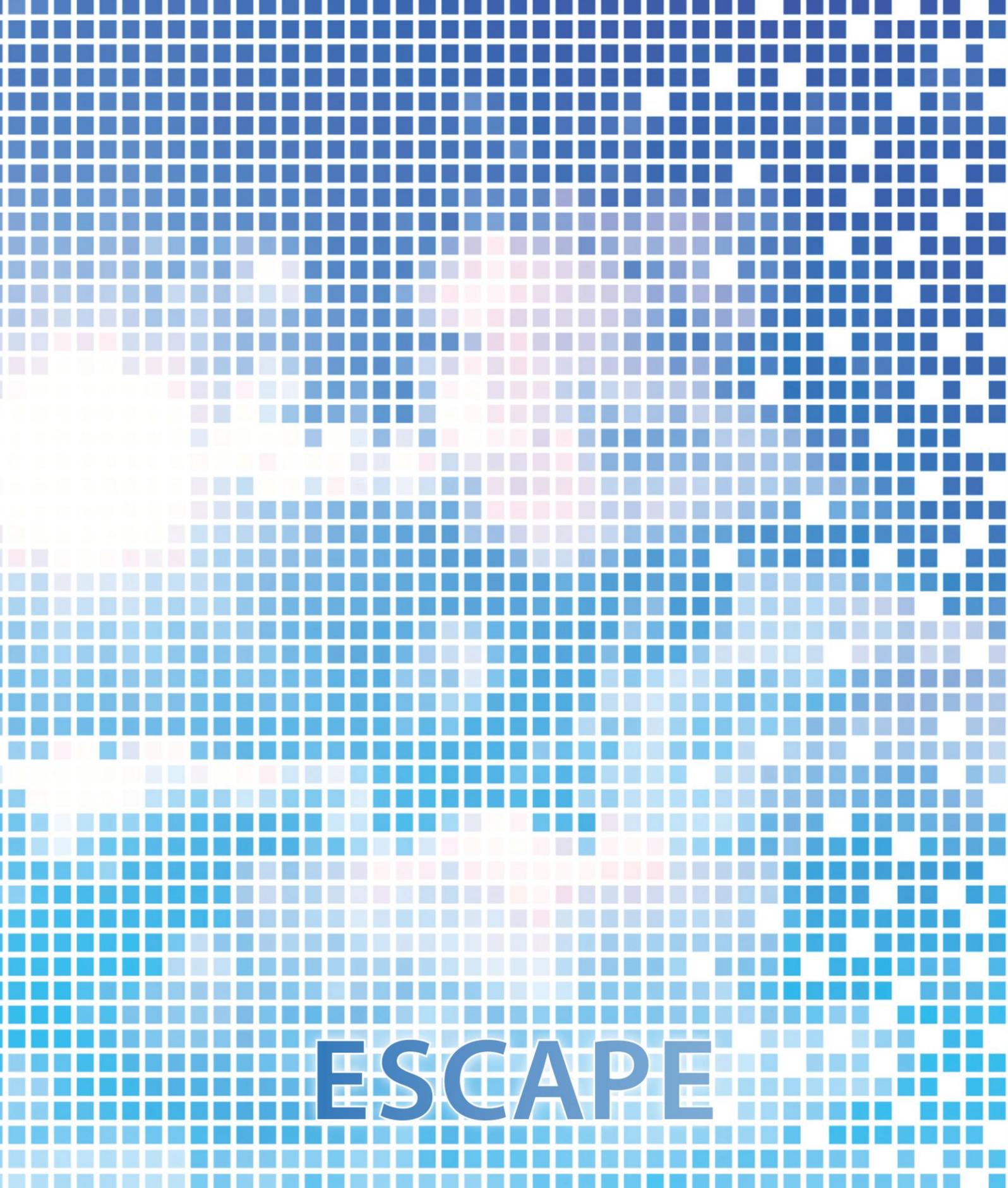